\documentclass[showpacs,10pt,twocolumn,prb]{revtex4-1}
\usepackage{amsmath}
\usepackage{amssymb}
\usepackage{graphics}
\usepackage{epsfig}
\usepackage{color}

\begin{document}

\title{Evolution of the Pauli spin-paramagnetic effect on the upper critical fields of K$_{x}$Fe$_{2-y}$Se$_{2-z}$S$_{z}$ single crystals}
\author{F. Wolff-Fabris,$^{1,\S}$ Hechang Lei,$^{2,\dag,\ddag}$ J. Wosnitza,$^{1,3}$ and C. Petrovic$^{2,\ddag}$}
\affiliation{$^{1}$Hochfeld-Magnetlabor Dresden (HLD), Helmholtz-Zentrum Dresden-Rossendorf, D-01314 Dresden, Germany\\
$^{2}$Condensed Matter Physics and Materials Science Department, Brookhaven
National Laboratory, Upton, New York 11973, USA\\
$^{3}$Institut f\"{u}r Festk\"{o}rperphysik, TU Dresden, D-01062, Dresden, Germany}
\date{\today}

\begin{abstract}
We have studied the temperature dependence of the upper critical fields,
$\mu_{0}H_{c2}$, of K$_{x}$Fe$_{2-y}$Se$_{2-z}$S$_{z}$ single crystals
up to 60 T. The $\mu _{0}H_{c2}$ for $H\parallel ab$ and $H\parallel c$ decrease
with increasing sulfur content. The detailed analysis using
Werthamer-Helfand-Hohenberg (WHH) theory including the Pauli
spin-paramagnetic effect shows that $\mu _{0}H_{c2}$ for $H\parallel ab$
is dominated by the spin-paramagnetic effect, which diminishes with higher
S content, whereas $\mu _{0}H_{c2}$ for $H\parallel c$ shows a linear
temperature dependence with an upturn at high fields. The latter
observation can be ascribed to multiband effects that become weaker for
higher S content. This results in an enhanced anisotropy of $\mu _{0}H_{c2}$
for high S content due to the different trends of the spin-paramagnetic
and multiband effect for $H\parallel ab$ and $H\parallel c$, respectively.
\end{abstract}

\maketitle

\section{Introduction}

Since the discovery of LaFeAsO$_{1-x}$F$_{x}$ with $T_{c}$ = 26 K,
\cite{Kamihara} there has been considerable effort invested in understanding
the properties and superconducting mechanism of
iron-based superconductors.\cite{Stewart,Dagotto,Hu} Thereby, the
temperature dependence of the upper critical field, $\mu_0H_{c2}$,
attracts great interests because it provides valuable information on
the coherence length, anisotropy, electronic structure, and
pair-breaking mechanism. However, iron-based superconductors exhibit
a rich diversity in the temperature dependence of $\mu_0H_{c2}$.
For FeAs-1111- and FeAs-122-type superconductors, such as La(O,F)FeAs
and Sr(Fe,Co)$_{2}$As$_{2}$,
the upper critical fields can be described using a two-band model.%
\cite{Hunte,Baily} For FeAs-111-
and FeSe-11-type
superconductors, such as LiFeAs and Fe(Te, Se/S), $\mu_0H_{c2}$ is
dominated by Pauli spin-paramagnetism.\cite{Khim,Lei1,Lei3}

Studies of the upper critical field in FeSe-122-type superconductors
($A_x$Fe$_{2-y}$Se$_2$, with $A =$ K, Rb, Cs, or Tl) are rare because
of the rather high superconducting transition temperature, $T_c$, and
concomitantly large zero-temperature critical field. In addition,
it is very challenging to handle the air-sensitive
samples. Mun et al. studied $\mu_0H_{c2}$ of K$_{0.8}$Fe$_{1.76}$Se$_{2}$
up to 60 T.\cite{Mun} They found that the upper critical field for
$H\parallel c$, $\mu _{0}H_{c2,c}$, increases quasilinearly with
decreasing temperature, whereas $\mu _{0}H_{c2,ab}$ for $H\parallel
ab$ flattens at low temperatures. The anisotropy of upper critical field,
$\gamma=H_{c2,ab}/H_{c2,c}$, decreases with $T$ and drops to about 2.5
at 18 K. A similar behavior has been observed for
Tl$_{0.58}$Rb$_{0.42}$Fe$_{1.72}$Se$_{2}$.\cite{Jiao} The analysis of
$\mu_0H_{c2}$ using the Werthamer-Helfand-Hohenberg (WHH) formula
including Pauli spin-paramagnetism and spin-orbit scattering indicates
that spin paramagnetism plays an important role in $\mu_0H_{c2,ab}$,
whereas the enhancement of $\mu_0H_{c2,c}$ at low temperatures is
likely attributed to multiband effects.\cite{Jiao}

In K$_{x}$Fe$_{2-y}$Se$_{2}$, substitution of Se by
sulfur suppresses $T_{c}$.\cite{Lei4} Preliminary measurements of
$\mu_0H_{c2}$ at low fields reveal that $\mu_0H_{c2}$ as well
decreases with S content, thereby showing temperature
dependence that can be described well using the simplified WHH model
without spin paramagnetism and spin-orbit scattering.\cite{Lei2} However,
the evolution of $\mu_0H_{c2}$ at very high fields and low temperatures
is still unclear. In this work, we report on the temperature dependence
and anisotropy of upper critical fields for three single crystals with
different S concentration, namely K$_{0.64}$Fe$_{1.44}$Se$_{2}$ (S-0),
K$_{0.70(7)}$Fe$_{1.55(7)}$Se$_{1.01(2)}$S$_{0.99(2)}$ (S-99), and
K$_{0.76(5)}$Fe$_{1.61(5)}$Se$_{0.96(4)}$S$_{1.04(5)}$ (S-104),
accessing pulsed magnetic fields up to 60 T. We found that the
spin-paramagnetic effect is rather important in $\mu_0H_{c2,ab}$,
but, at the same time, multiband effects dominate the temperature
dependence of $\mu_0H_{c2,c}$. These effects become
weaker with higher S content.

\section{Experiment}

The single crystals of K$_{x}$Fe$_{2-y}$Se$_{2-z}$S$_{z}$ used in this
study were grown and characterized as described previously.\cite{Lei4}
Magnetotrasport experiments in pulsed magnetic fields up to 62 T were
performed at the Dresden High Magnetic Field Laboratory facility, a
member of the European Magnetic Field Laboratory. Exposure of the
samples to ambient conditions was minimized by handling the samples
in a glove box. We have used standard four-contacts technique with
AC currents operating in the kHz frequency range. The electrical
resistance was measured by use of a fast
data-acquisition recording system and analyzed with a digital
lock-in technique. The contacts were made on freshly cleaved surfaces inside a glove box using silver paint and platinum wires. The contact resistance was between 10 and 50 Ohms and the excitation current was 0.3 mA which corresponds to the current density of approximately 10$^{3}$ A/m$^{2}$.  Anisotropic measurements were conducted on the same crystal.

K$_{x}$Fe$_{2-y}$Se$_{2}$ samples are intrinsically phase separated
into nanoscale magnetic insulating and Josephson-coupled superconducting
regions.\cite{Ryan,WangZ,LiuY,Ricci,LiW,YuanR,LoucaD,Lazarevic} The
insulating parts of the sample have several orders of magnitude higher
resistance ($R_{i}$) near $T_{c}$ when compared to the metallic parts
($R_{m}$).\cite{Shoemaker} Therefore, near $T_{c}$ and below
$R(T)\approx R_{m}(T)$. Recent angular resolved photoemission data
showed that the insulating parts of the sample do not contribute to
the spectral weight in the energy range near $E_{F}$,\cite{YiM} i.e.,
the temperature dependence of the resistivity below $T_{c}$
is dominated by the metallic parts of the sample,
similar to polycrystals having dense grain boundaries. Since sulfur substitution in K$_{x}$Fe$_{2-y}$Se$_{2}$ crystals is uniform,\cite{Lei4,HCLCC} S most likely substitutes both superconducting and insulating phase fractions.

\section{Results and Discussion}

\begin{figure}[tbp]
\centerline{\includegraphics[scale=0.42]{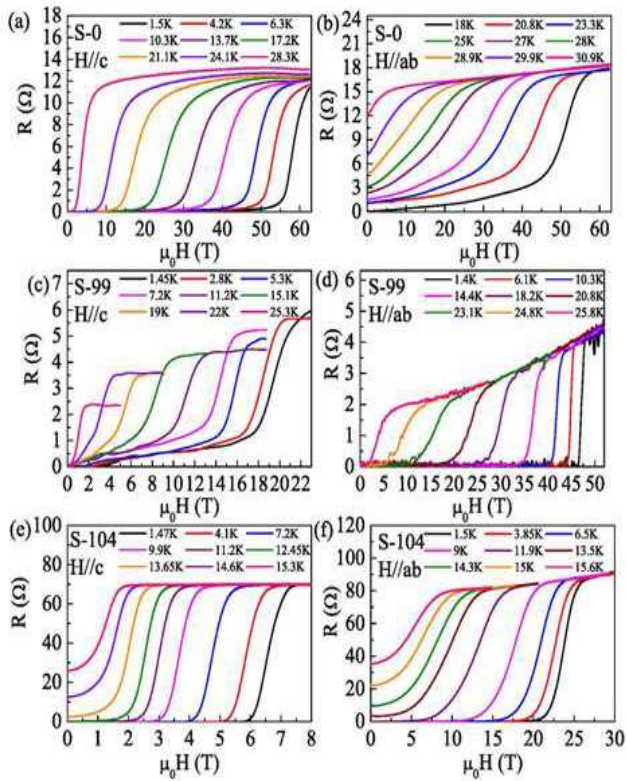}} \vspace*{-0.3cm}
\caption{(Color online) Magnetic-field dependence of the resistance,
$R$, of sample S-0 for (a) $H\parallel c$ and (b) $H\parallel ab$,
of sample S-99 for (c) $H\parallel c$ and (d) $H\parallel ab$, and
of sample S-104 for (e) $H\parallel c$ and (f) $H\parallel ab$ measured
at various temperatures.} \label{fig_1}
\end{figure}

Figure \ref{fig_1} shows the field dependence of the resistance, $R$, of the
samples S-0, S-99, and S-104 for $H\parallel c$ and $H\parallel ab$ at
various temperatures. Superconductivity is suppressed and the normal
state recovered with increasing magnetic fields at constant temperature
and the superconducting transitions in $R$ shift to lower magnetic
fields at higher temperatures. For some curves we observe a finite
resistance in the superconducting state that may be caused either by
experimental artifacts or by thermally activated
vortex-flux motion. The experimental artifacts may include a degradation
of the contacts or sample cracking during the course
of the experiment. For all samples, at the same temperature, the
transitions for $H\parallel ab$ occur at much higher fields when
compared to those for $H\parallel c$. This shows
that $\mu_0H_{c2,ab}$ is much larger than $\mu_0H_{c2,c}$ and that
there exists a large anisotropy in the upper critical fields for all
samples. On the other hand, for both field directions, the transitions
rapidly shift to lower fields with increasing S content. For example,
the superconducting transition at $T\sim$ 1.5 K changes from about 60
to 20 T and finally reaches 7 T for the samples S-0, S-99, and S-104,
respectively. This evidences that S doping significantly
suppresses $\mu_0H_{c2}$, consistent with previous results measured at
low fields.\cite{Ying,Lei2}

\begin{figure}[tbp]
\centerline{\includegraphics[scale=0.8]{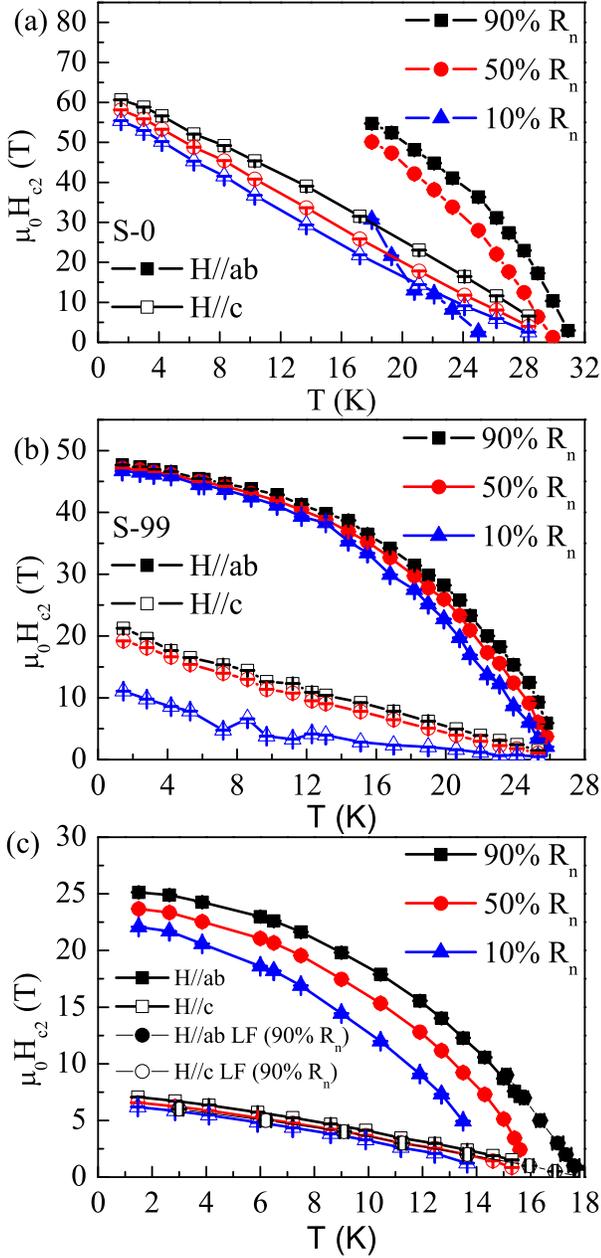}} \vspace*{-0.3cm}
\caption{(Color online) Temperature dependence of the resistive upper
critical fields, $\mu_0H_{c2}$ of the samples (a) S-0, (b) S-99, and
(c) S-104 for $H\parallel ab$ (closed symbols) and $H\parallel c$
(open symbols). Figure 2 includes low field data (LF) taken on S-104 sample used in Ref. 13 for comparison} \label{fig_2}
\end{figure}

Figure \ref{fig_2} presents the temperature dependence of the resistive upper
critical fields, $\mu_0H_{c2}$, of S-0, S-99, and S-104 determined from
the resistivity drops to 90\% (Onset), 50\% (Middle), and 10\% (Zero)
of the normal state resistance, $R_{n}$, for both field directions.
The normal-state resistance was determined by linearly extrapolating
the field-dependent resistance above the onset of the
superconductivity transition. The data taken in low fields are in good agreement for S-104 crystal [Fig. 2(c)]. The S-99 crystal had $T_{c}$ $\sim$ 26 K, somewhat higher than crystal used in low field studies,\cite{Lei2} but expected for crystals with that sulfur content.\cite{Lei4} For all samples, $\mu_0H_{c2}$ obtained
for $H\parallel ab$ is much larger than for $H\parallel c$, as mentioned
above. The temperature dependence of $\mu_0H_{c2,ab}$ for sample S-0
[Fig.\ \ref{fig_2}(a)] is distinctively different from that of
$\mu_0H_{c2,c}$. Close to $T_{c0}$ (zero-field transition
temperature), clearly different slopes are observed in the temperature
dependence of $\mu_0H_{c2}$ for both field orientations. With decreasing
temperature, the $\mu_0H_{c2,ab}$ curves start to bend downward with a
convex shape. In contrast, $\mu_0H_{c2,c}$ exhibits almost linear
temperature dependence. These results for the S-0 sample
are consistent with previous measurements using a contactless rf
technique.\cite{Mun,GasparovN} This suggests the absence of resistive heating in our measurements.

For the sample S-99 [Fig.\ \ref{fig_2}(b)], $\mu_0H_{c2}$ for both field
directions show a similar behavior as for S-0 but the absolute values
are much smaller than for the pure crystal. Moreover, $T_{c0}$ also
shifts to lower temperature. For sample S-104 [Fig.\ \ref{fig_2}(c)],
$\mu_0H_{c2,ab}$ and $\mu_0H_{c2,c}$ exhibit similar saturation trends
at low temperatures with different slopes at $T$ close to $T_{c0}$. When
compared to the previous results measured in low fields for sample S-104,
$\mu_0H_{c2,c}$ can be well described using the
simplified WHH formula, but $\mu_0H_{c2,ab}$ is remarkably smaller than
the value predicted.\cite{Lei2} This implies that the temperature
dependence of $\mu_0H_{c2}$ is influenced by factors outside the
simplified WHH model. Similarly, the nearly linear temperature
dependence of $\mu_0H_{c2,c}$ for S-0 and S-99 cannot be explained by
this model.

\begin{figure}[tbp]
\centerline{\includegraphics[scale=0.8]{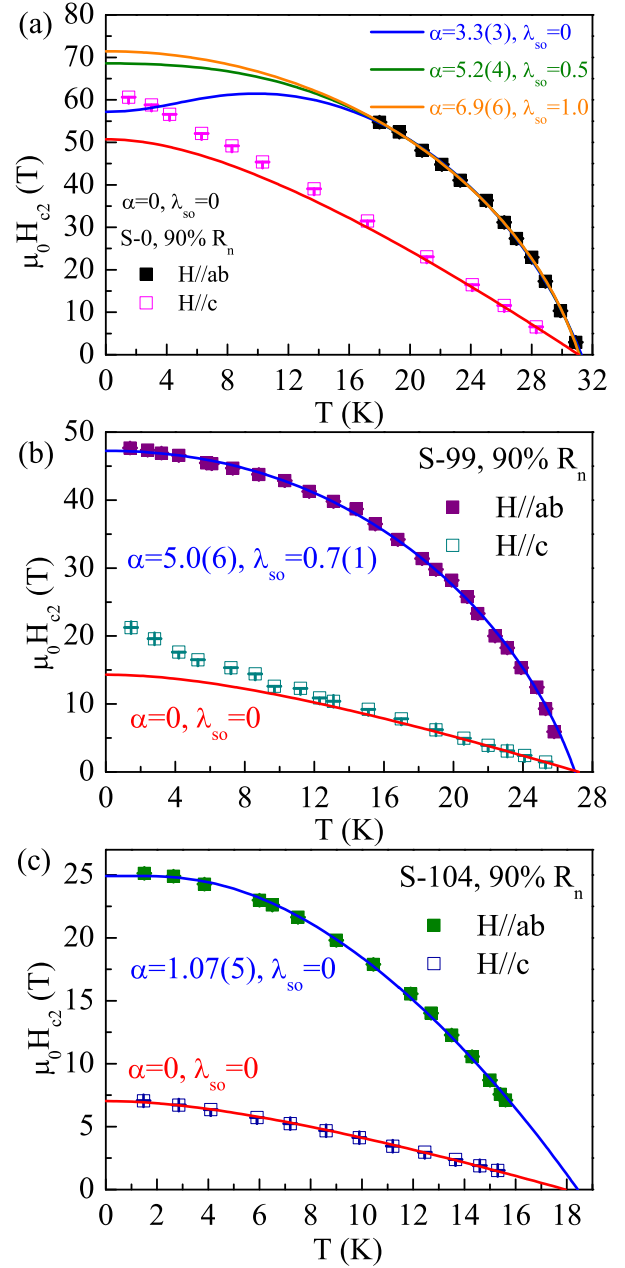}} \vspace*{-0.3cm}
\caption{(Color online) $\mu _{0}H_{c2}(T)$ determined from 90\% $R_{n}$ (symbols)
and fits using the WHH theory (solid lines) for the samples (a) S-0,
(b) S-99, and (b) S-104 at $H\parallel ab$ and $H\parallel c$.}
\label{fig_3}
\end{figure}

As previous studies have shown, spin paramagnetism has a significant
influence on the upper critical field of FeSe-11 and FeAs-122-type superconductors.%
\cite{Lei1,Lei3,Gasparov1,Gasparov2} We note that the Pauli paramagnetic limiting fields
in simplest approximation, $\mu_0H_{P}(0)=1.86T_{c}$,%
\cite{Chandrasekhar,Clogston} are about 58.0(2) T, 50.4(2) T, and
33.8(4) T for S-0, S-99, and S-104, respectively. These values are
comparable to the zero-temperature $\mu_0H_{c2,ab}$ for these samples.
This implies that the spin-paramagnetic effect needs to be considered
when analyzing the temperature dependence of $\mu_0H_{c2}$, especially
for $H\parallel ab$. Through the Maki parameter, $\alpha$, and
$\lambda_{so}$\cite{Maki} the effects of Pauli spin paramagnetism and
spin-orbit scattering have been included in the WHH theory for a
single-band $s$-wave weakly coupled type-II superconductor in the
dirty limit.\cite{Werthamer} $\mu_0H_{c2}$ is given by

\begin{equation}
\begin{array}{rcl}
\displaystyle\ln \frac{1}{t}=(\frac{1}{2}+\frac{i\lambda_{so}}{4\gamma})\psi(%
\frac{1}{2}+\frac{h+\lambda _{so}/2+i\gamma}{2t}) &  &  \\
\displaystyle+(\frac{1}{2}-\frac{i\lambda_{so}}{4\gamma})\psi (\frac{1}{2}+%
\frac{h+\lambda_{so}/2-i\gamma }{2t})-\psi(\frac{1}{2}), &  &
\end{array}%
\label{Eq_1}
\end{equation}

where $\psi(x)$ is the digamma function, $\gamma \equiv \lbrack (\alpha
h)^{2}-(\lambda _{so}/2)^{2}]^{1/2}$, and
\begin{equation}
h=\frac{4\mu _{0}H_{c2}(T)}{\pi ^{2}T_{c}(-d\mu _{0}H_{c2}(T)/dT)_{T=T_{c}}}.
\end{equation}

When $\alpha>1$, spin paramagnetism becomes essential.\cite{Maki}
In Fig.\ \ref{fig_3}, $\mu_0H_{c2}$ of S-0, S-99, and S-104, determined
using the 90\% $R_{n}$ criterion, together with the fits using formula
(\ref{Eq_1}) are shown. The 90\% data were chosen in
order to avoid the effects of flux motion and/or sample degradation.
When $\lambda_{so}$ is fixed to 0, $\mu_0H_{c2,ab}$ for sample S-0 can
be well described with $\alpha=$ 3.3(3) indicating
strong spin paramagnetism. It should be noted that the fit is not
unique and the temperature dependence of $\mu_0H_{c2,ab}$ can as well
be described with other fit values of $\alpha$ when $\lambda_{so}$
is non-zero. This is due to the limited temperature region where data
for $\mu_0H_{c2,ab}$ are available. For different
combinations of $\alpha$ and $\lambda_{so}$, $\mu_0H_{c2,ab}$
varies largely at low temperatures and high fields, below our
measurement range. Moreover, in order to describe
$\mu_0H_{c2}$ at low fields well, $\alpha$ has to increase
when $\lambda_{so}$ becomes larger because spin-orbit scattering tends
to reduce the effect of spin paramagnetism.\cite{Werthamer}
Accordingly, $\alpha=$ 3.3(3) is a lower limit. On the
other hand, the WHH model with $\alpha=$ 0 and $\lambda_{so}=$ 0 does
not describe the curve well. Thus, even though $\alpha$ and $\lambda_{so}$
are not uniquely determined by our data, the large $\alpha=$ 3.3(3)
strongly indicates that spin paramagnetism plays an important role in
suppressing superconductivity for $H\parallel ab$.
This sample might be worth to investigate in even
higher magnetic fields by using a 100 T coil in order to explore
the evolution of paramagnetic effects for $H\parallel ab$.

For $H\parallel c$, $\mu_0H_{c2,c}$ at low temperatures is enhanced
when compared to the WHH model. Such an enhancement
cannot be explained by $\alpha > 0$ or $\lambda_{so} > 0$. Indeed,
it suggests that multiband effect become important when the magnetic
field is oriented along the $c$ direction. A similar behavior has
been observed in Tl$_{0.58}$Rb$_{0.42}$Fe$_{1.72}$Se$_{2}$.\cite{Jiao}

Increasing the S content to 0.99, $\mu_0H_{c2,ab}$ is suppressed to
below 50 T in the whole temperature range [Fig.\ \ref{fig_3}(b)].
Without considering spin-orbital scattering, the WHH formula cannot
describe $\mu_0H_{c2,ab}$ well even when including spin paramagnetism.
When the spin-orbit scattering term is included, the fit quality
improves significantly and the obtained parameters are $\alpha=$ 5.0(6)
and $\lambda_{so}=$ 0.7(1). Assuming $\lambda_{so} = 0.7$ for the
sample S-0, $\alpha$ in sample S-0 will be larger than in
sample S-99. On the other hand, $\mu_0H_{c2,c}$ shows a linear
temperature dependence down to about 8 K, below which
an upturn appears. As for the pure sample, this upturn cannot be
described by use of the simplified WHH model. This shows that multiband
effects are still important in sample S-99 for $H\parallel c$.

For the single-band clean limit, the Maki parameters
for $H\parallel ab$ and $H\parallel c$ should be related by the ratio of
the Fermi velocity $v_{ab}$ in the $ab$ plane to $v_c$ along
$c$. The cylinder-like Fermi surface in iron-based
superconductors results in $v_{ab}\gg v_{c}$. Therefore, this ratio is
larger than 1 and the $\alpha$ for $H\parallel ab$ is expected to be
larger than for $H\parallel c$,\cite{Lei5} i.e., spin paramagnetism
is stronger for $H\parallel ab$ than for $H\parallel c$. The open
electronic orbits along $c$ reduce the orbital-limited upper critical
field considerably.\cite{Jiao}

With further S substitution [sample S-104, Fig.\ \ref{fig_3}(c)],
the WHH model without spin-orbit scattering can describe the
temperature dependence of $\mu_0H_{c2,ab}$ well with $\alpha = 1.07(5)$.
This is somewhat larger than 1, suggesting that
spin paramagnetism is still essential. But, the absolute value
is much smaller than for the samples S-0 and S-99. Obviously,
$\alpha$ decreases with increasing S content, i.e., spin paramagnetism
becomes less important. The $\mu_0H_{c2,c}$ can be well
described by using the WHH formula without including spin paramagnetism
or spin-orbit scattering. The slope $-d(\mu_0H_{c2,c})/dT|$ near
$T_{c0}$ is 0.565(5) T/K, which is very close to the previous results
measured at low fields.\cite{Lei2} Our results show that
the multiband as well as possible spin-paramagnetic effects are largely
suppressed in S-104 for $H\parallel c$.

\begin{figure}[tbp]
\centerline{\includegraphics[scale=0.75]{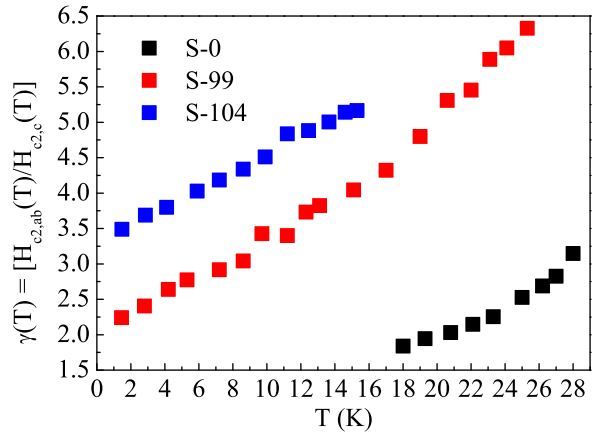}} \vspace*{-0.3cm}
\caption{(Color online) Temperature dependence of the anisotropy, $\gamma=H_{c2,ab}/
H_{c2,c}$ using the 90\% $R_{n}$ criterion for the samples S-0, S-99,
and S-104.}
\label{fig_4}
\end{figure}

The anisotropy of the upper critical field, $\gamma=H_{c2,ab}/H_{c2,c}$,
supplies further information on the effect of S substitution on the
evolution of $\mu_0H_{c2}$. As shown in Fig.\ \ref{fig_4}, $\gamma$
exhibits a similar trend in the temperature-dependence
for all three samples: the anisotropy decreases considerably with
temperature. For sample S-0, $\gamma$ is $\sim$ 3.1 at 28 K and
decreases to $\sim$ 1.8 at 18 K, a somewhat smaller value than reported
in literature.\cite{Mun} For sample S-99, $\gamma$ is about 6.3 at 25 K
and decreases gradually to $\sim$ 2.2 at 1.5 K. Finally, for sample
S-104, $\gamma$ lies between $\sim$ 5.2 at 15 K and $\sim$ 3.5 at 1.5 K.
This trend has been observed in all iron-based
superconductors.\cite{Jing,Lei5}

The origin of the small anisotropy of the upper critical fields
could be caused by a three-dimensional electronic structure,
spin paramagnetism, or multiband effects.\cite{Jing} The notable
outcome of our study is that the anisotropy
increases with S content, consistent with previous results measured
in the Ginzburg-Landau region where it was speculated that
this might be due to a smaller warping of the two-dimensional (2D)
Fermi surface (FS) with increasing S content.\cite{Lei2}
The increase of the anisotropy
with S substitution is understandable since orbital pair breaking is
more effective near $T_{c0}$,\cite{Jing} and a more 2D FS should lead
to a larger $\gamma$. However, this work shows that other factors
can as well result in an enhanced anisotropy of $\mu_0H_{c2}$ in
the high-field low-temperature region. First, spin paramagnetism
which usually decreases the anisotropy of the upper critical field
at high fields becomes weaker with increasing S content, thus the
suppression of $\mu_0H_{c2,ab}$ is reduced. Second, multiband effects
also become weaker at higher S content, and the slight upturn of
$\mu_0H_{c2,c}$ at high fields changes to a saturation behavior
described by the WHH model. These two opposite trends contribute
to a larger $\gamma$. Since $\alpha$ is proportional to the effective
mass in the clean limit,\cite{Gurevich1} the smaller $\alpha$ with
increasing S content can partially be related to a decrease of the
effective mass.\cite{Lei5,Wang} The decrease in $\alpha$ is in agreement with suppression of spin susceptibility and spin excitations in K$_{x}$Fe$_{2-y}$Se$_{1-z}$S$_{z}$.\cite{Torchetti} In addition, sulfur substitution
might also change the electronic structure leading to reduced
multiband effects. Further theoretical work is necessary to
clarify this.

\section{Conclusion}

In summary, we have investigated the upper critical fields of
K$_{x}$Fe$_{2-y}$Se$_{1-z}$S$_{z}$ single crystals up to 60 T.
The $\mu_0H_{c2}$ for both $H\parallel ab$ and $H\parallel c$ decreases
with increasing S content. For $H\parallel ab$, the $\mu_0H_{c2,ab}$
follows the WHH model including strong spin paramagnetism.
The $\mu_0H_{c2,c}$ for low S content shows a behavior that suggests
multiband effects and the single-band orbitally limited field
gradually becomes dominant at high sulfur content. The anisotropy
of $\mu_0H_{c2}$ is enhanced with increasing S content which can
be explained by weakened spin paramagnetism and reduced multiband
effects for $H\parallel ab$ and $H\parallel c$, respectively.

\section{Acknowledgements}

Work at Brookhaven is supported by the Center for Emergent
Superconductivity, an Energy Frontier Research Center funded
by the U.S. DOE, Office for Basic Energy Science (HL and CP).
Part of this work was supported by EuroMagNET II (EU Contract
No. 228043). CP acknowledges support by the Alexander von
Humboldt Foundation.

\S Current Address: European XFEL GmbH, Notkestrasse 85, 22607, Hamburg, Germany.

$\dag$ Current Address: Frontier Research Center and Materials and Structures Laboratory, Tokyo Institute of Technology, 4259 Nagatsuta, Midori, Yokohama 226-8503, Japan

\ddag\ hlei@lucid.msl.titech.ac.jp and petrovic@bnl.gov

\end{document}